\documentclass{article}
\usepackage{graphicx} 
\usepackage[margin=0.5in]{geometry}
\usepackage{placeins}
\usepackage{hyperref}
\usepackage{pythonhighlight}
\usepackage[authoryear]{natbib}
\usepackage{amsmath}
\usepackage{amsfonts}
\usepackage{titlesec}

\titlespacing*{\section}{0pt}{2em}{1em}

\titlespacing*{\subsection}{0pt}{3em}{1em}

\begin{document}

\begin{center}
\textbf{\Large Data-Driven Modeling for On-Demand Flow Prescription \\ in
    Fan-Array Wind Tunnels}

\vspace{2em}

{   Alejandro A. Stefan-Zavala$^{1\ast}$, 
    Isabel Scherl$^2$,
    Ioannis Mandralis$^1$,
    
    Steven L. Brunton$^3$,
    and Morteza Gharib$^1$}
\vspace{1em}
\end{center}

{\scriptsize
    \par$^1$\ Graduate Aerospace Laboratories, California Institute of Technology, Pasadena, CA 91125, USA
    
    \par$^2$\ Department of Mechanical and Civil Engineering, California Institute of Technology, Pasadena, CA 91125, USA
    
    \par$^3$\ Department of Mechanical Engineering, University of Washington, Seattle, WA 98195
    
$^\ast$\ Corresponding author. E-mail:
    \href{mailto:astefan@caltech.edu}{aastefan@caltech.edu}
}
\vspace{2em}
\hrule
\FloatBarrier
\vspace{1em}
\begin{centering}
{\large\textbf{Abstract}}\par\vspace{1em}
\hfill
\begin{minipage}{0.95\textwidth}
Fan-array wind tunnels are an emerging technology to design bespoke wind fields through grids of individually controllable fans.
This design is
especially suited for the turbulent, dynamic, non-uniform flow conditions found close to the ground, and has
enabled applications from entomology to flight on Mars. 
However, due to the high dimensionality of fan-array actuation and the complexity of unsteady fluid flow, the physics of fan arrays are not fully characterized, making it difficult to prescribe arbitrary flow fields.
Accessing the full capability of fan arrays requires resolving the map from time-varying grids of fan speeds to three-dimensional unsteady flow fields, which remains an open problem. This map is unfeasible to span in a single study, but it can be partitioned and studied in subsets.
In this paper, we study the special case of constant fan-speeds and 
time-averaged streamwise velocities with one homogeneous spanwise axis.
We~produce a proof-of-concept surrogate model by
fitting a regularized linear map to a dataset of fan-array measurements.
We use this model as the basis for an open-loop
control scheme to design flow profiles subject to constraints on fan speeds.
In experimental validation, our model scored a mean prediction error of 1.02 m/s and our control scheme a mean tracking error of 1.05 m/s in a fan array with velocities up to 12 m/s.
We empirically conclude that the physics relating constant fan speeds to 
time-averaged streamwise velocities are dominated by
linear dynamics, and present our method as
a foundational step to fully resolve fan-array wind tunnel control.
\end{minipage}
\hfill
\end{centering}


\newcommand{\Ndata}{N_\mathrm{data}}
\newcommand{\Ndatam}{{$\Ndata$\ }}

\newcommand{\Nfans}{{N_{\mathrm{fans}}}}
\newcommand{\Nfansm}{$\Nfans$\ }
\newcommand{\Nfansval}{10\ }
\newcommand{\Nsensors}{{N_{\mathrm{sensors}}}}
\newcommand{\Nsensorsval}{17\ }
\newcommand{\Ntrain}{169}
\newcommand{\Ntest}{63}
\newcommand{\ntest}{$\Ntest$ }
\newcommand{\Ntarget}{34}
\newcommand{\Niv}{34\ }
\newcommand{\Vmax}{11\ \text{m/s}}

\newcommand{\vvv}{\mathbf{v}}
\newcommand{\vvvm}{$\vvv$ }
\newcommand{\rr}{\mathbf{r}}
\newcommand{\rrm}{$\rr$ }
\newcommand{\riv}{\hat{\rr}}
\newcommand{\rivm}{$\riv$\ }
\newcommand{\R}{\mathbf{R}}
\newcommand{\Rm}{$\R$}
\newcommand{\V}{\mathbf{V}}
\newcommand{\Vm}{$\V$}
\newcommand{\A}{\mathbf{A}}
\newcommand{\Am}{$\A$ }
\newcommand{\bv}{{\mathbf{b}}}
\newcommand{\bvm}{$\bv$ }
\newcommand{\vpredm}{$\vpred$\ }

\newcommand{\vtrue}{\vvv_\mathrm{true}}
\newcommand{\vpred}{\vvv_\mathrm{pred}}
\newcommand{\mps}{\ \mathrm{m/s}}

\newcommand{\vtarget}{\vvv_\mathrm{target}}
\newcommand{\vtargetm}{$\vtarget$\ }
\newcommand{\vmeasured}{\vvv_\mathrm{measured}}
\newcommand{\mvmeasured}{$\vmeasured$\ }

\section{Introduction}
\label{sec:introduction}
\begin{figure}
    \centering
    \includegraphics[width=0.85\textwidth]{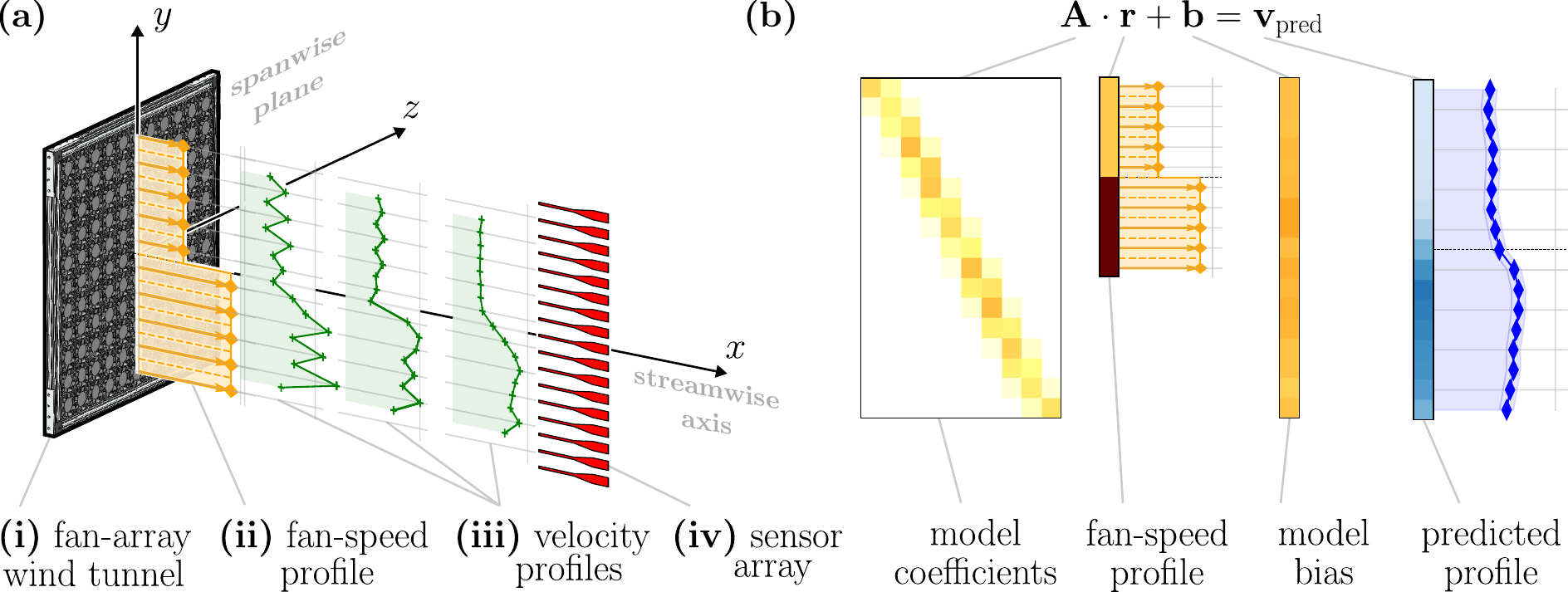}
    \caption{
    \textbf{(a) Left:} 
    a fan-array wind tunnel \textbf{(i)} is given static profiles of fan speeds 
     \textbf{(ii)} that vary along one spanwise axis ($y$).
     Mean profiles of streamwise velocity \textbf{(iii)}
     are measured by a sensor array \textbf{(iv)} along said axis ($y$).
    \textbf{(b) Right:} a vector of 
    fan speeds \rrm multiplied by a coefficient matrix 
    \Am, then added to a bias vector
    \bvm, gives a
    predicted velocity profile~ \vpredm.}
    \label{fig:intro}
\end{figure}

Fan-array wind tunnels (i.e. fan arrays, fan walls, multi-fan wind tunnels,
or Real Weather Wind Tunnels) provide an unprecedented ability to tailor fluid flows through the independent actuation of multiple fans. 
Compared to traditional wind tunnels with a single impeller, fan arrays have 
a smaller footprint for the same test-section size, have faster ramp times,
and produce slower, more turbulent flows (\cite{dougherty_experimental_2022}).
The most salient feature of fan arrays is the independent actuation of each fan. Partitioning the flow source into independent units
allows for spatial variation at the length-scale of one fan diameter
(\cite{dougherty_experimental_2022}).
The number of achievable \textit{static} flow profiles alone grows 
exponentially with the number of fans. Additionally, due to their small size and
mass, each fan can change speed dynamically throughout an experiment,
allowing for time-dependent flow variation.

The fan-array architecture has enabled novel wind tunnel research
across a wide range of applications.
\cite{veismann_low-density_2021} used the space
efficiency of fan arrays to conduct flight tests in a chamber
replicating the Martian atmosphere for the
\textit{Ingenuity} Mars Helicopter.
\cite{dougherty_experimental_2022} used the independent
actuation of 2,592 fans, arranged in a 3m $\times$ 3m grid, to
produce compound shear layers and 
``checkerboard" patterns in wind,
then varied
fan speeds sinusoidally to perturb shear-layer vortex-shedding frequencies,
in an effort to replicate
``flight-relevant environments."
\cite{oconnell_neural-fly_2022} used the same fan array, modulated 
sinusoidally, to produce dynamic wind conditions to 
develop, train and test the deep-learning autonomous
flight scheme \textit{Neural-Fly}. 
More examples of research leveraging fan arrays range from studies on
flies (\cite{lopez_characterization_2021}) to atmospheric turbulence 
(\cite{smith_simplified_2012},
\cite{ozono_realization_2018}
\cite{yos_improving_2019},
 \cite{veismann_characterization_2019}
 \cite{veismann_high_2020}, 
\cite{cui_generating_2021},
\cite{veismann_axial_2021}, 
\cite{walpen_real-scale_2023}, 
\cite{wei_phase-averaged_2022},
\cite{mokhtar_automated_2023},
\cite{dabiri_visual_2023}); to autonomous flight on Earth (\cite{olejnik_experimental_2022},  \cite{renn_applied_2023}, 
\cite{wang_turbulent_2021}); and on Mars (\cite{veismann_axial_2022}, \cite{veismann_axial_2021-1}).

%
The breadth of successful applications of fan arrays has motivated work on their flow characterization, physics modeling, and flow prescription
in recent years.
%
\cite{veismann_low-density_2021} determined a fan array's
functional test-section
(i.e. the sub-volume where turbulence intensity 
is minimal and boundary effects are negligible) when using only
a flush-mounted honeycomb for flow conditioning.
%
\cite{di_luca_design_2024} designed multi-layered flow management devices (FMD) to
improve fan-array flow quality by reducing turbulence intensity to a level seen in traditional, single-impeller wind tunnels (from $7\%$ to $0.45\%$) at the cost of reduced momentum output. 
%
\cite{dougherty_experimental_2022} 
showed that dual- and triple-stream free shear layers generated by
fan arrays follow the same self-similarity and scaling laws as
traditional ``splitter plate" shear layers (\cite{brown_density_1974}).
They also derived
an analytical model for the free-stream velocity of an array with
all fans set to the same time-varying speed. 
For this special case, they found that 
fan inertia and fluid inertia cause the fan array free-stream flow
to act as a low-pass filter of the fan-speed control signal.
\cite{di_luca_generation_2024} derived an analytical model for the streamwise velocity
of multi-plane, compound shear flows with augmented flow-conditioning.
They found that this case is governed by the velocity ratio between adjacent fans, consistent with \cite{dougherty_experimental_2022},
and that their model generalizes across downstream distances when normalized by
fan width.
%
%
\cite{walpen_automated_2024} used a proportional control scheme to prescribe flow
conditions enabled by feedback from a motion-tracked velocimetry system.
This scheme was validated by tracking a uniform flow target
of 8 m/s with mean absolute errors between 0.2 and 1~m/s.
Note how, across published works, fan width and total fan-array width
appear as length-scales of fan-array flows.


The aforementioned studies are highlights of an actively growing body of work,
each incrementally resolving fan-array physics and control.
However, due to the large fan-array state space and the complexity
of unsteady flow physics, designing arbitrary flows of the full
available dimensionality remains an open challenge and, thus,
the full capability of fan-array wind tunnels has not been achieved.
To achieve this ``full capability" means to prescribe both temporally and
spatially varying flow fields.
These flow fields can vary spatially along both spanwise axes of the
fan array, at a resolution of one fan width. They can also
vary temporally as fast as each fan can accelerate.

Tailoring flows in a fan array comes down to finding
the right fan speeds,
which can be done using a surrogate model.
%
A surrogate model predicts the behavior of the fan array
(given fan speeds, what flow is produced), and can be used to answer the
inverse question: given a target flow, what fan speeds produce it.
We call this process \textit{inverse design}.
An inverse-design scheme, once validated, can be used entirely
without sensor feedback, freeing the fan-array test-section;
or
it can be used within a feedback loop to converge in less time with
fewer sensors of lower resolution. 
A surrogate model that spans the full space of achievable
temporally and spatially varying flows 
is infeasible to resolve in a single study,
but it can be partitioned and systematically modeled in stages.

In this work, we focus on the subset of time-averaged streamwise velocity
profiles produced by steady fan speeds. 
We couple one axis of our fan array such that all rows of fans have the
same speed and only one spanwise axis is non-homogenous
(our flow profiles are ``one-dimensional").
See Figure~\ref{fig:intro}~(a).
For this special case, we produce a surrogate model and an open-loop scheme to
prescribe a desired flow.

We find that a regularized linear map is an effective surrogate model for this ``one-dimensional, static" fan-array case.
This linear map is encoded as a coefficient matrix and a bias vector, shown in Figure~\ref{fig:intro}~(b).
The resulting model coefficients match intuition: each velocity reading 
is primarily affected by the fan-row directly upstream of it,
secondarily affected by adjacent fan-rows, and insensitive to distant
fan-rows. Fitting models on profiles measured further
away from the source appears to capture the effect of viscous diffusion
and turbulent mixing, visible as a ``smearing" of model coefficients
(shown in Figure~\ref{fig:results_allmatrices}).

We apply this model as the basis of a linear program to
find the best fan speeds for desired velocity profiles via inverse design.
The linear program can be constrained on both fan speeds and 
predicted velocity profile. These constraints are critical in practice, where
noise restrictions, power limitations, and fan-unit failure can impose
new requirements on the state-space in which a model was trained.
We experimentally validate both out-of-sample prediction and
open-loop tracking performance using inputs and targets produced
after fitting. 

This paper is structured as follows: In {Section~\ref{sec:methods}}
we describe our 
fan array~(\ref{ssec:fawt}), sensors~(\ref{ssec:sensors}),
mathematics~(\ref{ssec:math}), and datasets~(\ref{ssec:dataset});
in {Section~\ref{sec:results}} we present the model
coefficients~(\ref{ssec:results_model}),
prediction performance~(\ref{ssec:results_prediction}),
and tracking performance~(\ref{ssec:results_tracking}) obtained
in validation experiments;
{Section~\ref{sec:conclusion}} contains concluding remarks
and considerations for 
future research towards complete fan-array modeling and control.

\section{Methods}\label{sec:methods}

\subsection{Experimental Design}
\label{ssec:experiment}

\begin{figure}
\includegraphics[width=0.95\linewidth]{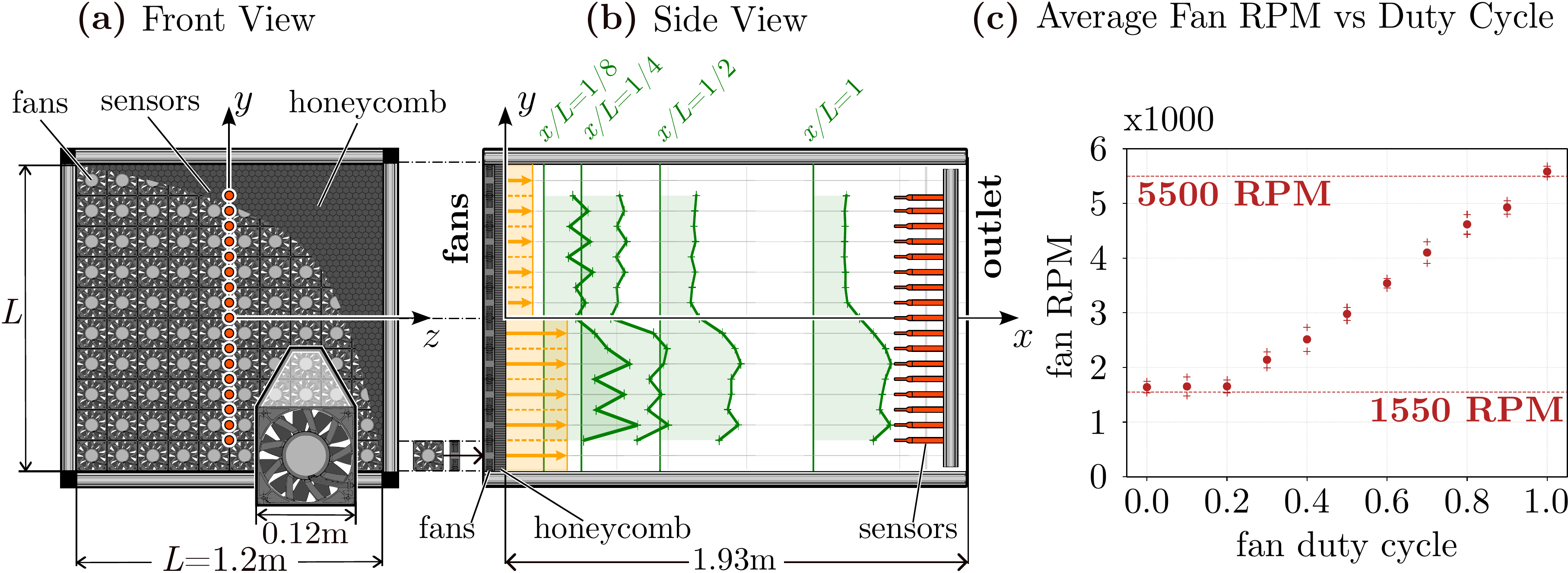}
  \caption{Experimental setup.
  \textbf{(a)}~Front view of $10\times 10$ fan array,
  showing the alignment of each sensor in orange circles
  along the middle of the array.
  \textbf{(b)}~Fan-array side view, with example velocity
  profiles at each of the four downstream distances
  measured. The sensor array is traversed
  along the $x$ axis.
  \textbf{(c)}~average RPM versus duty-cycle for all fans.
  }
\label{fig:geometry}
\end{figure}

\subsubsection{Fan Array Wind Tunnel}\label{ssec:fawt}
The model and control method presented in this paper are built entirely on experimentally collected fan array data.
There are three signals in a fan array flow experiment: 
(1) the control signal to each fan,
(2) the speed (rotational rate) of each fan and 
(3) the flow measured at some downstream location. 
Fan control signals are encoded as Pulse-Width Modulation (PWM) \textit{duty-cycle}, which ranges from $0$ or $0\%$
(lowest fan speed) to $1$ or $100\%$ (highest fan speed). 
Fan speed is measured in revolutions per minute (RPM).
Flow is measured in its streamwise component ($x$), is
averaged in time, and is
normalized by a reference velocity $V_\mathrm{max}=11\ \mathrm{m/s}$.

We use a $10\times 10$-fan array, 
composed of $100$ DELTA PFC1212DE-F00 $120\mathrm{mm}\times120\mathrm{mm}$ fan units.
This array spans a square $1.22\mathrm{m}\times1.22\mathrm{m}$ (spanwise) enclosed open-loop test section that extends $1.93\mathrm{m}$ downstream.
See Figure~\ref{fig:geometry}~(a)-(b). For this fan unit, the mapping from duty cycle to RPM is linear from $20\%$ duty cycle (nominally $1,550\ \mathrm{RPM}$) to $100\%$ duty cycle (nominally $5,500\ \mathrm{RPM}$). For duty cycles below $20\%$ these fans remain at their RPM floor of $1,550\ \mathrm{RPM}$. We use one flush-mounted $6.35\mathrm{mm}$ diameter, $38.1\mathrm{mm}$ thick ``honeycomb" for flow
conditioning. We normalize coordinates
in the test section by the width of the fan array, $L=1.2\mathrm{m}$. This is the same fan-array wind tunnel as used by
\cite{cosse_effect_2014}, \cite{sader_stability_2016}
and \cite{fan_effect_2019}. 

To control our fan array, we partition it into five $2\times10$ ``modules," each spanning two rows of the full $10\times10$ grid.
Each module has one NUCLEO F429ZI microcontroller, which directly controls and monitors each of its 20 fans.
All five microcontrollers are coordinated by a desktop computer over a Local Area Network.
Our fan-array control software is available open-source
(\cite{stefan-zavala_httpsgithubcomastefanzfan-club_2023}).

\subsubsection{Sensor Array}\label{ssec:sensors}
We use an array of \Nsensorsval Sensirion SDP31 digital differential pressure sensors arranged along the 
spanwise axis ($y$) and centered in the middle of our fan array.
The \Nsensorsval sensors are aligned with the centers and edges of fans.
(Figure~\ref{fig:geometry}~\textbf{(a)}-\textbf{(b)}).
The sensor array is traversed along the streamwise axis ($x$),
such that profiles are measured at 
different downstream distances within the $z=0$ plane.
The pressure sensors are placed in a 3D-printed housing 
and used as pitot-tube anemometers to measure streamwise
velocity.
%
All sensors are queried serially through the I2C serial protocol 
using an Adafruit~TCA9548A multiplexer.
A Teensy 4.0 microcontroller queries the full sensor array at
$9.2\ \mathrm{Hz}$. For a given fan-speed profile, data was collected for 7 seconds then averaged in time. 
%
Sensor readings are averaged in time and, at $x/L=1$,
have standard deviations
ranging from 0.06~$\mathrm{m/s}$ to 1.8~$\mathrm{m/s}$,
with an average standard deviation of 0.31~$\mathrm{m/s}$.
Our data collection and processing scripts are implemented using Jupyter Notebook 7.0 (\cite{kluyver_jupyter_2016}) and the Python 3.11.3 kernel,
as well as numpy version 1.26.4 (\cite{harris_array_2020}),
pandas version 2.2.1 (\cite{mckinney_data_2010}),
and Matplotlib version 3.8.4.

\subsubsection{Functional Test-Section: Fan Array Boundary Effects}
\label{sssec:test_section}

There are three key boundary effects for the test-section of a fan array:
close to the array, far from the array, and at the spanwise edges
of the array. These effects are studied in detail
by \cite{veismann_low-density_2021}
and \cite{dougherty_experimental_2022}.
When close to the fans (small $x/L$), the effect of fan geometry
(such as the size of a fan's hub and the boundary between each fan)
is observable in the flow profile.
With increasing distance from the fans (increasing $x/L$),
artifacts from fan geometry are dampened by viscous diffusion and
turbulent mixing. 
Diffusion and mixing, however,
also dampen variations due to nonuniform fan speeds, the
essential feature of fan arrays.
Measuring too close to the fan array introduces artifacts,
measuring too far negates individual-fan control.
For the fan array used in this paper, we found $x/L=1$ to be a good downstream distance based on this trade-off.

At the spanwise edges of the fan array
($|y/L|$ and $|z/L|$ close to $1/2$), a boundary layer will form
with the test-section enclosure (or a free-shear layer in the case
of an open test-section). \cite{veismann_low-density_2021}
recommends $y/L, z/L \in [-0.4, 0.4]$. The span of our sensor array
in this paper matches this recommendation.

\subsection{Mathematical Methods}\label{ssec:math}
\begin{figure}
\vspace{1em}
\centering
\begin{minipage}[t]{0.45\textwidth}
(a) \textbf{Surrogate Model}
\begin{python}
from sklearn.linear_model \
    import Lasso
# R.shape == (N_fans,N_data)
# V.shape == (N_sensors,N_data)
LS = Lasso(alpha=0.01)
LS.fit(R.T, V.T)
A = LS.coef_
b = LS.intercept_
# use: v_hat = A@r + b
#
# r.shape == (N_fans,)
# v.shape == (N_sensors,)
\end{python}
\end{minipage}
\hspace{1em}
\begin{minipage}[t]{0.45\textwidth}
(b) \textbf{Inverse Design}
\begin{python}
from scipy.optimize \
    import least_squares
# v_star.shape==(N_sensors,)
f = lambda r: (A@r+b)-v_star
# lb, ub are min,max fan dc
lb = np.zeros((N_fans,))
ub = np.ones((N_fans,))
r0 = np.random.rand(N_fans)
result = least_squares(f,r0,
    bounds=(lb, ub),
    loss='soft_l1')
r_hat = result.x
\end{python}
\end{minipage}
\caption{Implementation of mathematical methods in Python 3. \textbf{(a)} Regression of surrogate model given input dataset \Rm\ and output dataset $\V$. \textbf{(b)} Inverse-design application of the surrogate model to estimate the best input \rivm for a desired profile $\vtarget$.}
\label{fig:implementation}
\vspace{0.5em}
\end{figure}

\subsubsection{Surrogate Model}
\label{sssec:model}

Let $\rr=[r_1,\ r_2,\ ...\ r_\Nfans]$ be a fan-speed profile.
The component $r_j$ is the duty cycle of 
the $j^\mathrm{th}$ actuator, and is a real-valued 
scalar ranging from $0$ (lowest fan speed) to $1$ (maximum fan speed).
Since we couple each row of fans in our $10\times 10$ array 
to the same duty cycle, we have $\Nfans=\Nfansval$ actuators and $r_j$ is the
$j^{\mathrm{th}}$ \textit{row} of fans (Figure~\ref{fig:geometry}~(a)).
We say \rrm is an input into the fan array.

Let $\vvv = [v_1,\ v_2,\ ...\ v_\Nsensors]$ be a velocity profile at a fixed 
downstream distance.
The component $v_i$ is the
nondimensionalized, 
time-averaged, streamwise velocity reading of the $i^\mathrm{th}$
sensor in an array of $\Nsensors$ sensors, and is a real-valued positive scalar.
In this paper, we have $\Nsensors=\Nsensorsval$
sensors along the nonhomogeneous spanwise axis $y$
(Figure~\ref{fig:geometry}~(b)).
We say \vvvm is an output of the fan array.
A fan-array surrogate model predicts
the output \vvvm produced by a given input~\rm.

Let $N$ be the number of all experimentally collected profiles
at a given downstream distance,
each consisting of an 
input-output pair $(\rr, \vvv)$. Let
$\R\in\mathbb{R}^{n\times N}$ be the matrix of $N$ 
horizontally stacked input vectors and
 $\vvv\in\mathbb{V}^{m\times N}$ be the matrix
of $N$ horizontally stacked output vectors, 
$$
\mathbf{R} = \begin{bmatrix}
    |     &     |     &     |     &     &       |       \\
\rr_1 & \rr_2 & \rr_3 & ... & \rr_{N} \\
    |     &     |     &     |     &     &       |       \\
\end{bmatrix} \quad
\mathbf{V} = \begin{bmatrix}
    |     &     |     &     |     &     &       |       \\
\vvv_1 & \vvv_2 & \vvv_3 & ... & \vvv_{N} \\
    |     &     |     &     |     &     &       |       \\
\end{bmatrix}.
$$

Our surrogate model consists of an $m\times n$ coefficient matrix \Am
and an $m$-dimensional bias vector \bvm.
Given an input \rm, an estimate \vpredm of 
the true output \vvvm is given by
$$
\vpred = \A\rr + \bv.
$$

We produce our surrogate model $(\A, \bv)$ from our data matrices
\Rm\ and \vvvm\ using regularized linear regression. This means we 
find the \Am and \bvm that best map \Rm\ to \vvvm such that $\A\R + \bv \approx \vvv$.
This is the same as minimizing
the $l_2$ norm $||\A\R + \bv - \vvv||_2$.
To regularize, we penalize the $l_1$ norm
(the sum of coefficient magnitudes) of \Am and \bvm.
The explicit form of our regression is

$$
\A,\bv = \underset{{\A,\bv}}{\text{argmin}}\ \  \frac{1}{2\Ndata   }||\A\R + \bv - \V||^2_2 + \alpha (||\A||_1 + ||\bv||_1).
$$

For our final model, we chose a regularization parameter $\alpha=0.01$.
Our implementation in code is given in Figure~\ref{fig:implementation}~(a).
It is built on Scikit-learn 1.5.1 (\cite{pedregosa_scikit-learn_2011}).

Linear regression regularized by the $l_1$ norm is known as LASSO
regression.
Regularizing on the $l_1$ norm promotes model sparsity,
which is a proxy for simplicity.
The effect of regularization on our model
is discussed in Section~\ref{ssec:results_model}.


\begin{figure}
    \centering
    \includegraphics[width=1\textwidth]{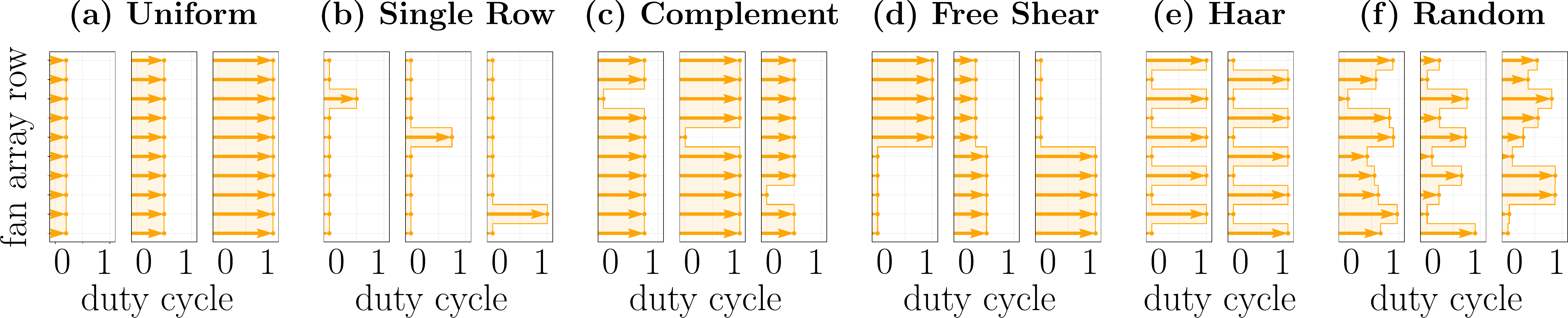}
    \caption{Types of inputs used in data collection.
    \textbf{(a)~Uniform}: all fans to the same duty cycle.
    \textbf{(b)~Single Row:} one fan-row at nonzero
    duty-cycle, all other fans at zero duty-cycle.
    \textbf{(c)~Complement:} one fan-row at zero
    duty-cycle, all others at nonzero duty-cycle.
    \textbf{(d)~Free Shear:} top half and bottom
    half of the fan array at different duty cycles.
    \textbf{(e)~Haar:} Concatenated Haar wavelets
        at different resolutions, and their mirror profiles.
    \textbf{(f)~Random:} each fan-row at a random
    duty-cycle, sampled from a uniform distribution.}
    \label{fig:inputs}
\end{figure}

\subsubsection{Inverse Design}
\label{ssec:inverse-design}

Given a desired flow profile \vtargetm,
we use our surrogate model $(\A,\bv)$ to find the input \rivm
that produces the closest profile to $\vtarget$:
$$
\begin{aligned}
\hat{\rr}=\min_{\rr} & ||\A\rr + \bv - \vtarget ||_2 \\
\mathrm{subject\ to} & \\
& l_k \leq r_k \leq u_k \\
\end{aligned}
$$
where $l_k$ and $u_k$ are lower and upper bounds on the duty-cycle of each fan.
Use them to encode limitations on fan speed,
such as saturation ($l_k=0, u_k=1$) or `dead' fans ($l_k=u_k=0$).
Our implementation in code is given in Figure~\ref{fig:implementation}~(b).
It is built on SciPy 1.12.0 (\cite{virtanen_scipy_2020}).

\subsection{Data Description}
\label{ssec:dataset}

To fit our models, we start with a set 
$\mathbf{R}_\mathrm{train}$ of \Ntrain\ inputs.
For each input, we measure the output
profile at four downstream distances:
$x/L=1/8$, $x/L=1/4$, $x/L=1/2$ and $x/L=1$, using
the setup described in Section~\ref{ssec:experiment}.
For each downstream distance, we compile a 
set $\mathbf{V}_\mathrm{train}$ of \Ntrain\ 
measured outputs corresponding to
$\mathbf{R}_\mathrm{train}$.
For each pair
$(\mathbf{R}_\mathrm{train},
\mathbf{V}_\mathrm{train})$
we fit a surrogate model $(\A, \b)$ as described in
Section~\ref{sssec:model}.

The inputs that form set $\mathbf{R}_\mathrm{train}$
fall within six ``types" of profiles, 
described in Figure~\ref{fig:inputs}.
These profile types were chosen for their
relevance in previous fan-array experiments
(Uniform and Free-Shear),
their potential use as basis for 
the fan-array input space (Single-Row and Complement), 
and for broad coverage of the input space
(Haar and Random).
%
The 169 profiles of the training set
break down into types as follows: 
11 Uniform,
60 Single-Row,
40 Complement,
10 Free Shear,
8 Haar,
and
40 Random.
The resulting models are 
shown and described in Section~\ref{ssec:results_model}.

To validate our model predictions, we use a
set $\mathbf{R}_\mathrm{test}$ of \Ntest\ 
inputs. Test inputs are of the same profile types
as training inputs, but with different parameters and
duty-cycles. At each modeled downstream distance
a set of predicted profiles
$\hat{\mathbf{V}}_\mathrm{test}$ is 
compared against a set of measured profiles
$\mathbf{V}_\mathrm{test}$.
Prediction performance is described in
Section~\ref{ssec:results_prediction}

We validate the tracking performance of our
inverse-design scheme using a set of \Ntarget\
target \textit{output} profiles (not inputs).
Tracking performance is described in
Section~\ref{ssec:results_tracking}

\section{Results}
\label{sec:results}

\subsection{Surrogate Model}
\label{ssec:results_model}

\newcommand{\restrim}{3.81cm 4.2cm 3cm 1cm}
\newcommand{\pht}{10em}
\begin{figure}
    \centering

    \includegraphics[width=0.75\textwidth]{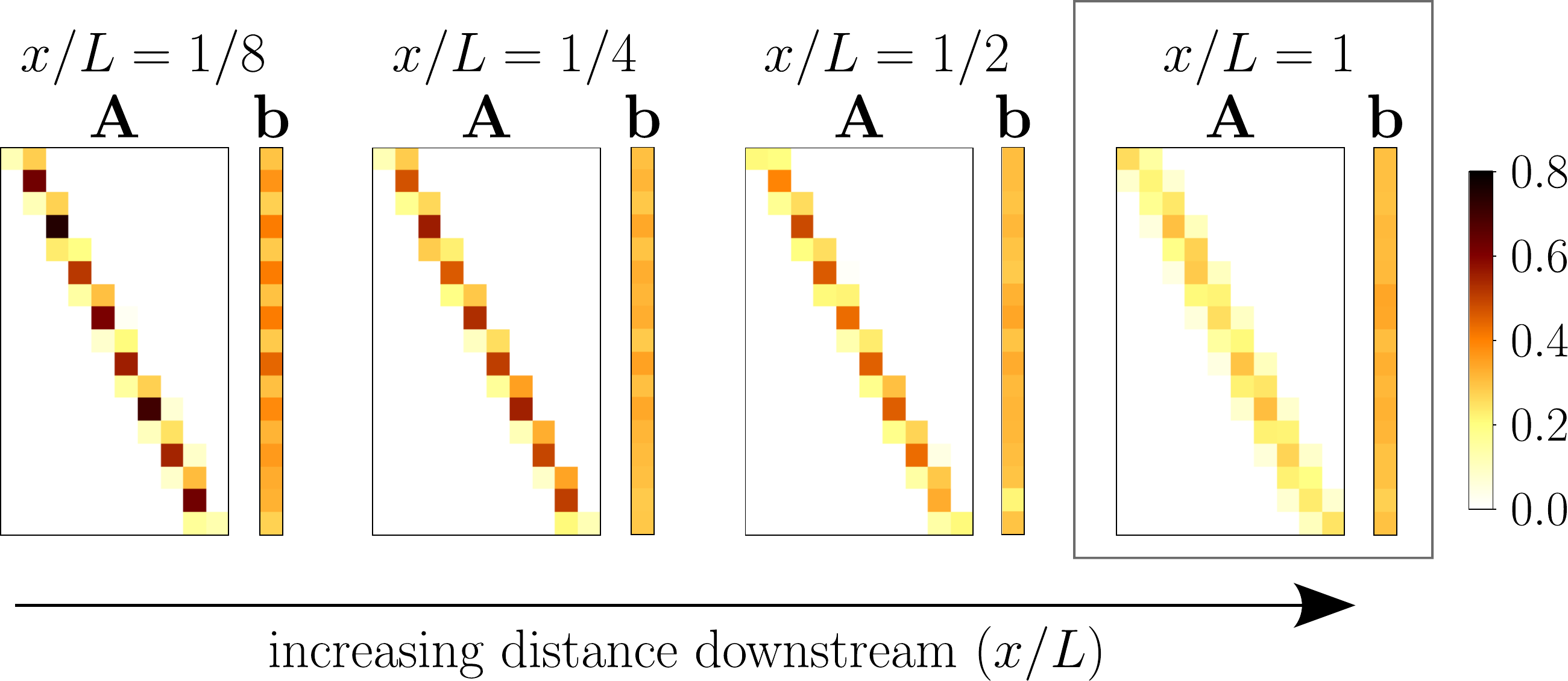}
    
    \caption{
    %
    Resulting model coefficient matrices ($\A$)
        and bias vectors ($\bv$).
%
    %
    As $x/L$ increases, model coefficients appear to ``diffuse."
 %
    To the right, highlighted, $x/L=1$ is the
    preferred distance for the fan array modeled in this study.}
    
    \label{fig:results_allmatrices}
\end{figure}
%
%
%
The surrogate model $(\A, \bv)$ (Section
\ref{sssec:model}) was fit to map between training
dataset pairs $(\R, \vvv)$ at the four measured locations $x/L\in\{1/8, 1/4,1/2,1\}$ (Section \ref{ssec:dataset}).
The resulting model coefficients and biases are shown in 
  Figure \ref{fig:results_allmatrices}.
According to the model coefficients, each sensor reading
(row of $\A$) is primarily
affected by the speed of fans (columns of $\A$)
directly upstream,
secondarily affected by adjacent fans, and 
insensitive to all other fans. This is encoded by the 
tri-diagonal-like structure of $\A$.
Comparing model coefficients as $x/L$ increases
shows effects that may be attributed to  
turbulent mixing and viscous diffusion as flow structures advect downstream. That is, the influence of each fan mixes with that
of neighboring fans:
As $x/L$ increases, model coefficients become less sparse,
smaller in magnitude, and more similar.
The influence of fan geometry is evident when the coefficients
for sensors aligned with fan \textit{hubs} are different
from those aligned with fan \textit{edges}.
This is the case for  all measured $x/L$ \textit{less than} 1.
where coefficients at fan hubs
(near 0.6 for $x/L=1/8$) are up to three times
larger than 
coefficients for fan edges (near 0.2 for $x/L=1/8$).
At $x/L=1$, 
the coefficients at fan hubs
and fan edges are most similar, ranging from
0.1 to 0.25. Therefore, at $x/L=1$ 
the effect of fan geometry is small,
while custom velocity profiles are still achievable.
It is from this trade-off between velocity profile 
smoothness and controllability that
$x/L=1$ is the preferred downstream distance
for this fan array (Section \ref{sssec:test_section}).
%
Bias vectors $\bv$ (Section~\ref{sssec:model}) capture the fixed nonzero fan RPM
for duty cycles at or below 0.3
specific to this fan array
(Section~\ref{ssec:fawt}, Figure~\ref{fig:geometry}~\textbf{(c)}),
which results in a fixed velocity
``floor" of 0.3$V_\mathrm{max}$ or 0.34 m/s at $x/L=1$.

The further away we measure from the source,
the more linear the map between fan and velocity
profiles appears to be, and therefore the
better our models fit. 
%
%
Similarly, the average $R^2$
value across all \Nsensorsval output variables 
is lowest at $x/L=1/8$ with $R^2=0.74$,
highest at $x/L=1/2$ with $R^2=0.81$, and
is  $R^2=0.79$ at $x/L=1$.
These $R^2$ values suggest
most of the variation in sensor readings is explained by
the linear fits, and moreso with increasing $x/L$.

\newcommand{\fh}{10em}
\subsection{Output Prediction}
\label{ssec:results_prediction}
\begin{figure}
    \centering
 \includegraphics[width=0.8\textwidth]{
         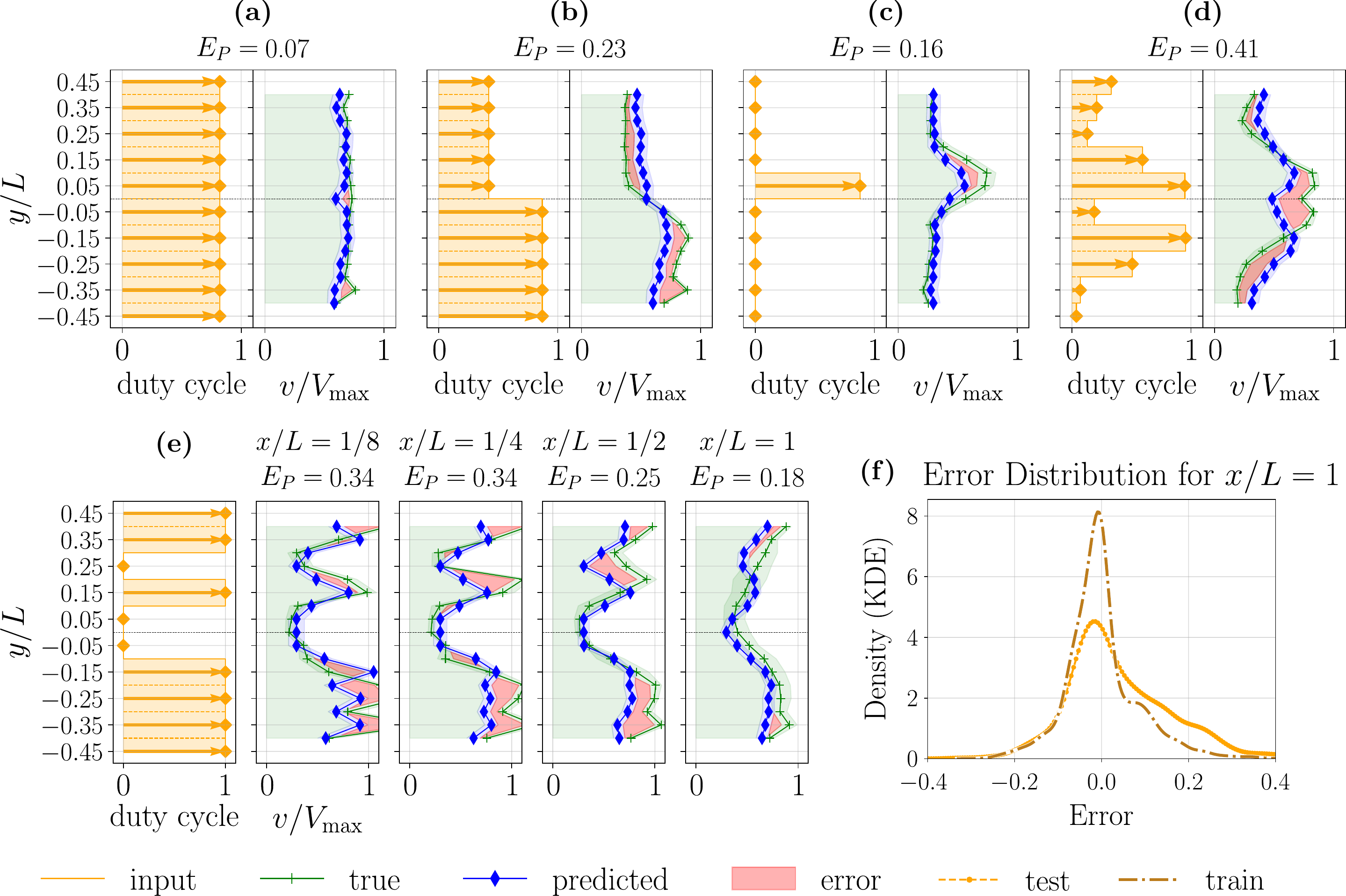}
    \caption{
    \textbf{Top.}
    Example profiles from test-set
    at $x/L=1$, showing input profile,
    model prediction and true measurement.
    \textbf{Bottom.}
    \textbf{Left:}~Example test-set profiles at
    all modeled downstream distances.
    \textbf{Right:}~Kernel-Density plots
    showing the
    distribution of component-wise
    (signed) prediction errors ($v-\hat{v}$)
    for both train and test set.
    Sample-wise mean absolute percent error (MAPE) $E_P$ is
    shown above each sample.
    }
    \label{fig:results_prediction}
\end{figure}

Out-of-sample (not seen in training)
performance was tested with  predictions for
\ntest brand new fan array inputs versus their measured
``true" velocity profiles at each $x/L$.
We quantify \textit{prediction error} as
mean absolute error (MAE) between measured and predicted
profiles.
For a \textit{single} input $\rr$,
prediction $\vpred = \A\rr + \bv$ and
true measurement $\vtrue$,
the prediction error of this one sample is the
average absolute value of the component-wise difference between $\vtrue - \vpred$:
\begin{equation}
\label{eqn:mae}
\mathrm{MAE}(\vtrue, \vpred) = 
\frac{1}{N_\mathrm{sensors}}\sum_k^{N_\mathrm{sensors}} |v_k - \hat{v}_k|
    \quad\quad \begin{matrix}
    \vtrue = [v_1,v_2,...v_k,...v_{N_\mathrm{sensors}}]^\mathrm{T} \\
    \vpred = [\hat{v}_1,\hat{v}_2,...\hat{v}_k, ...\hat{v}_{N_\mathrm{sensors}}]^\mathrm{T}
    \end{matrix}
\end{equation}

The prediction error for an entire dataset is the average of all sample-wise MAEs. Since our profiles are normalized by
$V_\mathrm{max}=\Vmax$, this MAE is in nondimensional velocity
as a fraction of $V_\mathrm{max}$.
Multiplying by $V_\mathrm{max}$ gives the 
dimensional MAE's reported in this section.

At the preferred downstream distance $x/L=1$, testing
MAE is 0.093 $v_\mathrm{error}/V_\mathrm{max}$ or $1.02\mps$
and training MAE is 0.057 $v_\mathrm{error}/V_\mathrm{max}$ or 0.63 m/s.
This performance is consistent across downstream distances:
The additional distances $x/L\in\{1/2,1/4,1/8\}$ have
dimensional testing MAE's of
1.04 m/s, 1.06 m/s and 1.16 m/s, respectively,
and dimensional training MAE's of 
0.61 m/s, 0.72 m/s and 0.86 m/s, respectively.
This gives an average out-of-sample (test-set)
prediction error
of 1.07 m/s across all measured $x/L$.

We additionally compute mean absolute percentage error
(MAPE), which normalizes the component-wise error 
by the absolute value of the $\vtrue$
component:
\begin{equation}
\label{eqn:mape}
\mathrm{MAPE}(\vtrue, \vpred) = 
\frac{1}{N_\mathrm{sensors}}\sum_k^{N_\mathrm{sensors}} \frac{|v_k - \hat{v}_k|}{\max(|v_k|, \epsilon)}
    \quad \begin{matrix}
    \vtrue = [v_1,v_2,...v_k,...v_{N_\mathrm{sensors}}]^\mathrm{T} \\
    \vpred = [\hat{v}_1,\hat{v}_2,...\hat{v}_k, ...\hat{v}_{N_\mathrm{sensors}}]^\mathrm{T}
    \end{matrix}
\end{equation}



$\epsilon$ is an arbitrary, small constant used to
prevent division by zero.
At the preferred downstream distance $x/L=1$,
training and testing MAPE's are 0.132 (13\%)
and 0.173 (17.3\%), respectively.
The additional distances, $x/L\in\{1/8, 1/4, 1/2\}$
have training MAPE's of 0.158, 0.146, 0.126, 
and testing MAPE's of 0.186, 0.176, 0.164, respectively.
The MAPE's of individual samples are shown in 
Figure~\ref{fig:results_prediction}.
MAE and MAPE are calculated using the Scikit-learn~1.5.1
implementations
\texttt{mean\_absolute\_error} and
\texttt{mean\_absolute\_percentage\_error},
respectively
 (\cite{pedregosa_scikit-learn_2011}).

Kernel-density plots of signed, component-wise error distributions
($v_\mathrm{true}-v_\mathrm{pred}$) at $x/L=1$, shown in
Figure~\ref{fig:results_prediction}~\textbf{(f)},
show a bias towards positive error or undershooting
($v_\mathrm{pred} < v_\mathrm{true}$), stronger in testing than in training.
This pattern is observable in all samples shown in 
Figure~\ref{fig:results_prediction}~\textbf{(a)}-\textbf{(e)},
where most errors (red shade) consist of measured profiles (green crosses)
being larger (to the right) of predicted profiles (blue diamonds).
We attribute most error to nonlinear fluid dynamic effects, which
a linear map of unmodified input features cannot capture. 




\subsection{Inverse Design}\label{ssec:results_tracking}
\renewcommand{\fh}{8em}

\begin{figure}
    \centering
\includegraphics[width=0.8\textwidth]{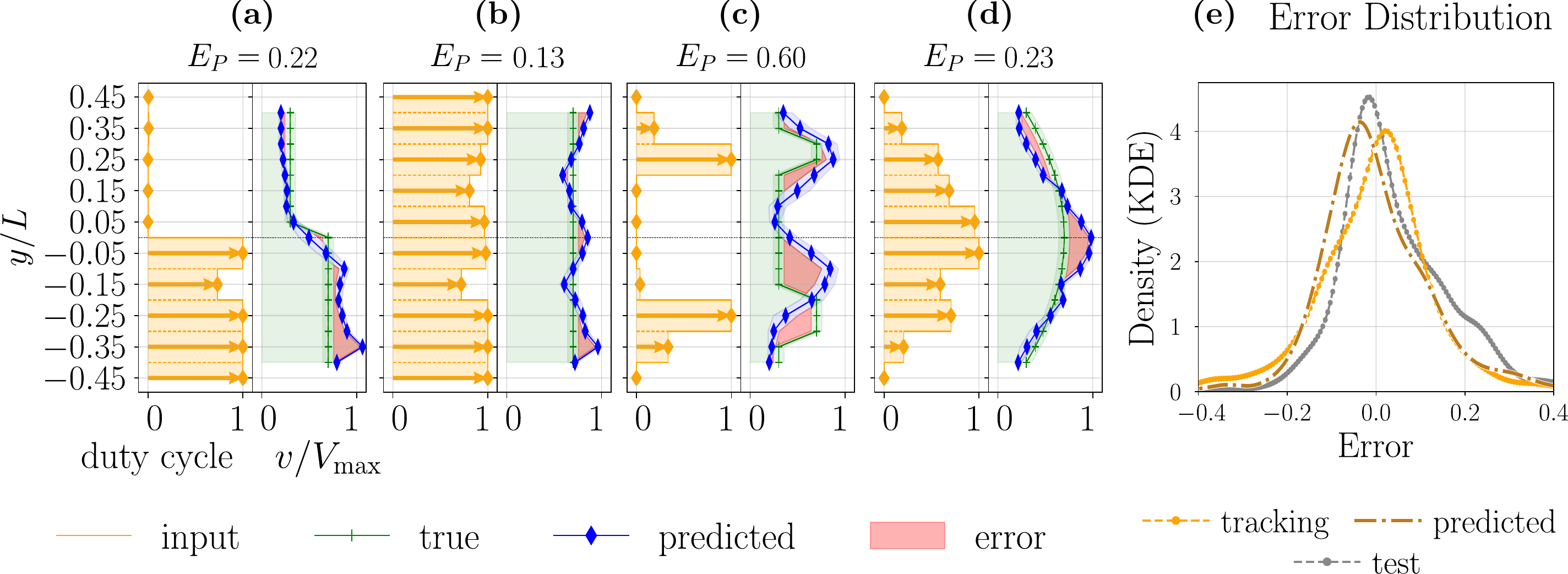}
\caption{Example profiles from inverse-design
validation at $x/L=1$,
showing resulting control input
(yellow), target flow (green) and true
measurement (blue).
\textbf{Rightmost:}~Kernel-Density plot of the
distribution of (signed) component-wise
tracking error ($v_\mathrm{target} - v$), along with
prediction error \textit{for inverse-design
inputs}, and prediction test-error from
Figure~\ref{fig:results_prediction}.
Sample-wise mean absolute percent error (MAPE) $E_P$ is 
shown above each sample.
}
\label{fig:results_tracking}
\end{figure}







Tracking performance of the inverse-design open-loop control
algorithm described in Section~\ref{ssec:inverse-design} was
tested at the preferred downstream distance
$x/L=1$ using a set of \Niv target velocity profiles not present 
in testing or training datasets.

We quantify \textit{tracking error} the same way as prediction error
in Section~\ref{ssec:results_prediction}, as mean absolute error
between a \textit{target} velocity profile \vtargetm and the
\textit{measured} velocity profile \mvmeasured that results
from executing the fan array input \rrm produced by the
inverse-design algorithm to track \vtargetm, the tracking error
of a single sample is given by $\mathrm{MAE}(\vtarget,\vmeasured)$ in
\ref{eqn:mae}. The tracking error of the entire inverse-design
test-dataset is the average of all sample-wise tracking errors.
The $x/L=1$ inverse-design test-set has a 
normalized MAE of
0.095 $v_\mathrm{error}/V_\mathrm{max}$ and
a MAPE of 0.215. This gives an
average dimensional tracking error of 1.05~m/s for $x/L=1$.
Test-set prediction MAE for $x/L=1$ is 1.02~m/s
(Section~\ref{ssec:results_prediction}). 
MAE suggests the inverse-design open-loop control 
algorithm is
as effective at tracking target profiles as the underlying linear model is at predicting profiles,
if slightly worse at lower velocities, as suggested by
higher MAPE.
This consistency in performance between tracking and prediction is
visible in
Figure~\ref{fig:results_tracking}~\textbf{(e)}.

Comparing the distribution of signed tracking errors in
Figure~\ref{fig:results_tracking}~\textbf{(e)} shows complementary
biases: prediction error is biased towards underestimating
outputs and tracking error is, in turn, biased towards overshooting targets.
The profiles in
Figure~\ref{fig:results_tracking}~\textbf{(a)-(d)} show this
overshooting, where measured profiles (blue diamonds) are most often
of larger magnitude (to the right) of target profiles (green crosses).
Additionally, as is the case with prediction error,
the dominant source of error appears to be
nonlinear fluid-dynamic effects, most evident in Figure~\ref{fig:results_tracking}~\textbf{(c)-(d)}.

\section{Conclusion}
\label{sec:conclusion}

The key feature of fan arrays is the independent control
of multiple flow sources. 
Space efficiency, rapid speed-modulation and
characteristic turbulence
are also distinguishing, useful features
worthy of study. 
However, independent control of each impeller is
the feature that sets fan arrays apart as a potential
field of study of their own: the field of replicating
arbitrary fluid-flows by finding suitable fan speeds.
This fine control is also the feature with the most complexity
and unanswered questions.
At its full richness, using a fan array means 
prescribing a three-component, unsteady turbulent velocity field
in a three-dimensional volume, as a function of time-varying grids.
%

In this work, we modeled the steady-state, streamwise
velocity profiles of an array of coupled rows of fans,
as a function of fan duty-cycle.
In search of the simplest model that works, 
we used $l_1$-regularized linear regression to encode
steady-state fan array physics as a sparse coefficient matrix
and a bias vector. 
The parsimony of our design resulted in a model that is not
only simple and effective, but also interpretable.
The matrix coefficients show an intuitive relation between
sensors and actuators. Comparing the model fit at
different downstream distances shows the effect of
viscous diffusion and turbulent mixing 
(Figure~\ref{fig:results_allmatrices}).
We successfully applied and validated this model,
both to predict the output of out-of-sample
fan-array inputs and to find desired velocity
profiles subject to constraints on fan speeds.
Though restricted to a special case,
our model covers a
large and vastly applicable range of fluid flows. 
Steady-state streamwise velocity
profiles that vary along one axis
are a large swath of classical fluid dynamics.
We can now prescribe shear layers without
splitter plates and
Couette flows without moving walls, and we can
change shear ratios and forcing velocities 
without modifying the test section.
We can even prescribe some unsteadiness.
For example, although the profile 
of a shear layer is steady, the Kelvin-Helmholtz
instability at its core has rich turbulent
dynamics which are a function of the steady-state
profile (\cite{dougherty_experimental_2022}).
Finally, the open-loop nature of our inverse-design
scheme allows for flow prescription without a
sensor array occupying the test section,
where one typically places the object of
study in an experiment.

One avenue for future work in this area is to better resolve the model under the present assumptions. 
In this work, our model was fit from fan duty-cycles
to flow velocities.
Though there is a linear relation between fan duty-cycle
and RPM (Figure~\ref{fig:geometry}-\textbf{(c)}), fitting a map
from RPM to flow velocity captures the underlying physics
more directly than from duty-cycle to flow velocity.
Duty-cycle is the control signal into a fan, while RPM is the
actual rotational speed of the fan.
Fans of the same model have slight variations in RPM for the 
same duty cycle. A fan that is blocked, damaged or worn out will
also have its RPM response affected.
A map from RPM to flow velocity will hold true even as 
fan units deteriorate, are perturbed, or are replaced.
Though only one spanwise direction was resolved,
we hypothesize that our exact method (Section~\ref{sssec:model}),
can be applied to a setting with both spanwise axes actuated
(i.e. fan-rows decoupled and the full fan-array is controllable).
Our inverse-design scheme is effective without feedback. However, 
this model can also be used as the basis for closed-loop control.
A feedback control scheme would compensate for unmodeled dynamics,
a dominant source of error in
Sections~\ref{ssec:results_prediction} and
~\ref{ssec:results_tracking}.
Looking at the coefficient matrices in Section~\ref{ssec:results_model}
suggests an underlying smooth kernel, which is superimposed 
at the time of prediction.
The alternating magnitudes of coefficients with
fan hubs and fan edges resembles superimposing Gaussian
profiles centered at each fan.
Finding this kernel may enable a resolution-invariant,
self-similar representation of the fan array, where prediction
is done by convolving this kernel with a curve of 
fan speeds.
As shown in Section~\ref{ssec:results_model},
our linear fit captured most of the underlying pattern 
in our training data,
with an $R^2$ value around 0.8 across all datasets.
The remaining unmodeled dynamics can be captured by
adding nonlinear input features.
The added model complexity can be
kept minimal with regularization that promotes sparsity,
such as the present $l_1$ regularization, or with a more
sophisticated method that enables a large library of candidate
nonlinearities (\cite{brunton_discovering_2016}).

Another direction for continued research is to relax simplifying assumptions and model new regimes of fan-array dynamics. 
One such extension is time-resolved fan-array flows.
Modeling flows from inputs that both vary in time
(e.g. varying fan-speeds in sine waves as done by
\cite{ozono_realization_2018}
and \cite{dougherty_experimental_2022})
and are resolved along a spanwise axis,
would enable the prescription of arbitrary
unsteady flow profiles.
A similar approach can be used for frequency content.
The modes of a flow can be added as input features and
modeled as a function of fan speeds,
which makes flow power spectrum prescribable.
Controlling power spectrum would
allow for power-spectra matching of flows measured in nature,
and for tailored studies of frequency response in
experiments.
Another two dimensions to model are the spanwise components
of the velocity field. In fan arrays where all fan outlets point
in the same direction, flow will be dominated by its streamwise
component. Still, there exist applications in which spanwise
velocities are critical, such as spanwise gusts that
perturb autonomous fliers from the side or from below.

As shown in Figure~\ref{fig:results_allmatrices}, there is a clear
pattern of diffusion and smearing in model coefficients with
increasing distance downstream.
If this diffusion effect is modeled,
one can obtain coefficients for all locations in the test section
despite having only measured at a few streamwise locations.
Finally, a large enough body of work would allow for the map
from fan specifications and array configuration to the
surrogate model that would be obtained from such a design.
Predicting what surrogate model will be obtained by
a fan-array enables the highest abstraction of inverse-design:
given the flows you wish to make, what fan array should you build and how should you control it.
This would attain the most general and resolved level of
fan-array modeling and control.
With this capability,
any gust, breeze, plume or tide, from anywhere on Earth or Mars,
can be replicated, faithfully, in lab, as can all the
phenomena of Nature and engineering that
fly, swim, float or exist therein.



\paragraph{Acknowledgements}
A.S.Z. Acknowledges
Amrita Mayavaram and Peter Renn for the development of 
    the anemometers used in data collection;
Maya Madere for the wiring and assembly of the initial sensor
    array,
Noel Esparza-Duran for the reassembly and restoration of
    the first and second fan arrays used;
Chris Dougherty, Marcel Veismann and David Kremers 
    for their mentorship and initial conceptualization of 
    fan-array modeling and control;
Peter Crocker and Steve Madere for discussions leading to
    instrumental insights;
as well as family, friends, and colleagues -- especially 
    Samm Crocker, Jack Caldwell, Julian Humml, Michael Ol and 
    Ariane Mora -- for their support.
The figures in this manuscript were produced using Matplotlib (\cite{hunter_matplotlib_2007}).


\paragraph{Funding Statement}
S.L.B. Acknowledges funding support from the Air Force Office of Scientific Research (FA9550-21-1-0178).

\paragraph{Declaration of Interests}
The authors declare no conflict of interest.

\paragraph{Author Contributions}
conceptualization: A.S.Z., S.L.B., I.M.;
data collection -- proof-of-concept: A.S.Z., I.M.;
data collection -- for publication: A.S.Z., I.S.;
formal analysis: A.S.Z., I.S., S.L.B.;
visualization: A.S.Z., I.S., S.L.B.;
funding acquisition: M.G.;
supervision: I.S., S.L.B., M.G.;
writing -- original draft: A.S.Z;
writing -- review and editing: A.S.Z., I.S., S.L.B., I.M.

\paragraph{Data Availability Statement}
Raw data are available from the corresponding author
A.S.Z., upon reasonable request.

\paragraph{Ethical Standards}
The research meets all ethical guidelines, including
adherence to the legal requirements of the study country.

\paragraph{Supplementary Material}
Code supplement available at \url{https://github.com/astefanz/fml-1}.


\bibliographystyle{apalike}
{\small\bibliography{references}}

\begin{thebibliography}{}

\bibitem[Brown and Roshko, 1974]{brown_density_1974}
Brown, G.~L. and Roshko, A. (1974).
\newblock On density effects and large structure in turbulent mixing layers.
\newblock {\em Journal of Fluid Mechanics}, 64(4):775--816.

\bibitem[Brunton et~al., 2016]{brunton_discovering_2016}
Brunton, S.~L., Proctor, J.~L., and Kutz, J.~N. (2016).
\newblock Discovering governing equations from data by sparse identification of nonlinear dynamical systems.
\newblock {\em Proceedings of the National Academy of Sciences}, 113(15):3932--3937.
\newblock Publisher: Proceedings of the National Academy of Sciences.

\bibitem[Cossé et~al., 2014]{cosse_effect_2014}
Cossé, J., Sader, J., Kim, D., Cerdeira, C.~H., and Gharib, M. (2014).
\newblock The {Effect} of {Aspect} {Ratio} and {Angle} of {Attack} on the {Transition} {Regions} of the {Inverted} {Flag} {Instability}.
\newblock American Society of Mechanical Engineers Digital Collection.

\bibitem[Cui et~al., 2021]{cui_generating_2021}
Cui, W., Zhao, L., Cao, S., and Ge, Y. (2021).
\newblock Generating unconventional wind flow in an actively controlled multi-fan wind tunnel.
\newblock {\em Wind and Structures}, 33(2):115--122.
\newblock Number: 2.

\bibitem[Dabiri et~al., 2023]{dabiri_visual_2023}
Dabiri, J.~O., Howland, M.~F., Fu, M.~K., and Goldshmid, R.~H. (2023).
\newblock Visual anemometry for physics-informed inference of wind.
\newblock {\em Nature Reviews Physics}, 5(10):597--611.
\newblock Publisher: Nature Publishing Group.

\bibitem[Di~Luca et~al., 2024a]{di_luca_design_2024}
Di~Luca, M., Leipold, M., and Noca, F. (2024a).
\newblock Design, implementation and validation of a flow management device for fan-array wind tunnels.
\newblock In {\em {AIAA} {SCITECH} 2024 {Forum}}. American Institute of Aeronautics and Astronautics.
\newblock \_eprint: https://arc.aiaa.org/doi/pdf/10.2514/6.2024-2673.

\bibitem[Di~Luca et~al., 2024b]{di_luca_generation_2024}
Di~Luca, M., Leipold, M., and Noca, F. (2024b).
\newblock On the generation of plane compound shear flows with fan-array wind tunnels.
\newblock In {\em {AIAA} {SCITECH} 2024 {Forum}}. American Institute of Aeronautics and Astronautics.
\newblock \_eprint: https://arc.aiaa.org/doi/pdf/10.2514/6.2024-2672.

\bibitem[Dougherty, 2022]{dougherty_experimental_2022}
Dougherty, C.~J. (2022).
\newblock {\em On the {Experimental} {Simulation} of {Atmospheric}-{Like} {Disturbances} {Near} the {Surface}}.
\newblock phd, California Institute of Technology.

\bibitem[Fan et~al., 2019]{fan_effect_2019}
Fan, B., Huertas-Cerdeira, C., Cossé, J., Sader, J.~E., and Gharib, M. (2019).
\newblock Effect of morphology on the large-amplitude flapping dynamics of an inverted flag in a uniform flow.
\newblock {\em Journal of Fluid Mechanics}, 874:526--547.
\newblock Publisher: Cambridge University Press.

\bibitem[Harris et~al., 2020]{harris_array_2020}
Harris, C.~R., Millman, K.~J., van~der Walt, S.~J., Gommers, R., Virtanen, P., Cournapeau, D., Wieser, E., Taylor, J., Berg, S., Smith, N.~J., Kern, R., Picus, M., Hoyer, S., van Kerkwijk, M.~H., Brett, M., Haldane, A., del Río, J.~F., Wiebe, M., Peterson, P., Gérard-Marchant, P., Sheppard, K., Reddy, T., Weckesser, W., Abbasi, H., Gohlke, C., and Oliphant, T.~E. (2020).
\newblock Array programming with {NumPy}.
\newblock {\em Nature}, 585(7825):357--362.
\newblock Publisher: Nature Publishing Group.

\bibitem[Hunter, 2007]{hunter_matplotlib_2007}
Hunter, J.~D. (2007).
\newblock Matplotlib: {A} {2D} {Graphics} {Environment}.
\newblock {\em Computing in Science \& Engineering}, 9(3):90--95.
\newblock Conference Name: Computing in Science \& Engineering.

\bibitem[Kluyver et~al., 2016]{kluyver_jupyter_2016}
Kluyver, T., Ragan-Kelley, B., Pérez, F., Bussonnier, M., Frederic, J., Hamrick, J., Grout, J., Corlay, S., Ivanov, P., Abdalla, S., and Willing, C. (2016).
\newblock Jupyter {Notebooks}—a publishing format for reproducible computational workflows.

\bibitem[Lopez, 2021]{lopez_characterization_2021}
Lopez, A. (2021).
\newblock Characterization of fluid flow inside a multi-fan-array wind tunnel for insects.
\newblock page F20.002.
\newblock Conference Name: APS Division of Fluid Dynamics Meeting Abstracts ADS Bibcode: 2021APS..DFDF20002L.

\bibitem[McKinney, 2010]{mckinney_data_2010}
McKinney, W. (2010).
\newblock Data {Structures} for {Statistical} {Computing} in {Python}.
\newblock {\em Proceedings of the 9th Python in Science Conference}, pages 56--61.
\newblock Conference Name: Proceedings of the 9th Python in Science Conference.

\bibitem[Mokhtar et~al., 2023]{mokhtar_automated_2023}
Mokhtar, N.~O., Fernández-Cabán, P.~L., and Catarelli, R.~A. (2023).
\newblock Automated large-scale and terrain-induced turbulence modulation of atmospheric surface layer flows in a large wind tunnel.
\newblock {\em Experiments in Fluids}, 65(1):5.

\bibitem[O'Connell et~al., 2022]{oconnell_neural-fly_2022}
O'Connell, M., Shi, G., Shi, X., Azizzadenesheli, K., Anandkumar, A., Yue, Y., and Chung, S.-J. (2022).
\newblock Neural-{Fly} enables rapid learning for agile flight in strong winds {\textbar} {Science} {Robotics}.

\bibitem[Olejnik et~al., 2022]{olejnik_experimental_2022}
Olejnik, D.~A., Wang, S., Dupeyroux, J., Stroobants, S., Karasek, M., De~Wagter, C., and de~Croon, G. (2022).
\newblock An {Experimental} {Study} of {Wind} {Resistance} and {Power} {Consumption} in {MAVs} with a {Low}-{Speed} {Multi}-{Fan} {Wind} {System}.
\newblock In {\em 2022 {International} {Conference} on {Robotics} and {Automation} ({ICRA})}, pages 2989--2994.

\bibitem[Ozono and Ikeda, 2018]{ozono_realization_2018}
Ozono, S. and Ikeda, H. (2018).
\newblock Realization of both high-intensity and large-scale turbulence using a multi-fan wind tunnel.
\newblock {\em Experiments in Fluids}, 59(12):187.

\bibitem[Pedregosa et~al., 2011]{pedregosa_scikit-learn_2011}
Pedregosa, F., Varoquaux, G., Gramfort, A., Michel, V., Thirion, B., Grisel, O., Blondel, M., Prettenhofer, P., Weiss, R., Dubourg, V., Vanderplas, J., Passos, A., Cournapeau, D., Brucher, M., Perrot, M., and Duchesnay, A. (2011).
\newblock Scikit-learn: {Machine} {Learning} in {Python}.
\newblock {\em Journal of Machine Learning Research}, 12(85):2825--2830.

\bibitem[Renn, 2023]{renn_applied_2023}
Renn, P. I.~J. (2023).
\newblock {\em Applied {Machine} {Learning} for {Prediction} and {Control} of {Fluid} {Flows}}.
\newblock phd, California Institute of Technology.

\bibitem[Sader et~al., 2016]{sader_stability_2016}
Sader, J.~E., Huertas-Cerdeira, C., and Gharib, M. (2016).
\newblock Stability of slender inverted flags and rods in uniform steady flow.
\newblock {\em Journal of Fluid Mechanics}, 809:873--894.

\bibitem[Smith et~al., 2012]{smith_simplified_2012}
Smith, J.~T., Masters, F.~J., Liu, Z., and Reinhold, T.~A. (2012).
\newblock A simplified approach to simulate prescribed boundary layer flow conditions in a multiple controlled fan wind tunnel.
\newblock {\em Journal of Wind Engineering and Industrial Aerodynamics}, 109:79--88.

\bibitem[Stefan-Zavala, 2023]{stefan-zavala_httpsgithubcomastefanzfan-club_2023}
Stefan-Zavala, A.~A. (2023).
\newblock https://github.com/astefanz/fan-club.
\newblock original-date: 2023-05-10T18:48:59Z.

\bibitem[Veismann, 2022]{veismann_axial_2022}
Veismann, M. (2022).
\newblock {\em Axial {Descent} of {Multirotor} {Configurations} -- {Experimental} {Studies} for {Terrestrial} and {Extraterrestrial} {Applications}}.
\newblock phd, California Institute of Technology.

\bibitem[Veismann et~al., 2021a]{veismann_axial_2021}
Veismann, M., Burdick, J., Gharib, M., Wei, S., Conley, S., Young, L., Delaune, J., and Izraelevitz, J. (2021a).
\newblock Axial {Descent} of {Variable}-{Pitch} {Multirotor} {Configurations}: {An} {Experimental} and {Computational} {Study} for {Mars} {Deployment} {Applications}.
\newblock The Vertical Flight Society.

\bibitem[Veismann et~al., 2021b]{veismann_low-density_2021}
Veismann, M., Dougherty, C., Rabinovitch, J., Quon, A., and Gharib, M. (2021b).
\newblock Low-density multi-fan wind tunnel design and testing for the {Ingenuity} {Mars} {Helicopter}.
\newblock {\em Experiments in Fluids}, 62(9):193.

\bibitem[Veismann and Gharib, 2019]{veismann_characterization_2019}
Veismann, M. and Gharib, M. (2019).
\newblock Characterization of {Rotor}-{Rotor} {Aerodynamic} {Interactions} for {Free} {Flight} {Studies} of {Multirotor} {Systems}.
\newblock page B17.003.
\newblock Conference Name: APS Division of Fluid Dynamics Meeting Abstracts ADS Bibcode: 2019APS..DFDB17003V.

\bibitem[Veismann and Gharib, 2020]{veismann_high_2020}
Veismann, M. and Gharib, M. (2020).
\newblock High {Fidelity} {Aerodynamic} {Force} {Estimation} for {Multirotor} {Crafts} in {Free} {Flight}.
\newblock In {\em {AIAA} {Scitech} 2020 {Forum}}, {AIAA} {SciTech} {Forum}. American Institute of Aeronautics and Astronautics.

\bibitem[Veismann et~al., 2021c]{veismann_axial_2021-1}
Veismann, M., Wei, S., Conley, S., Young, L., Delaune, J., Burdick, J., Gharib, M., and Izraelevitz, J. (2021c).
\newblock Axial {Descent} of {Variable}-{Pitch} {Multirotor} {Configurations}: {An} {Experimental} {Investigation} under {Free}-{Flight} {Settings} for {Mars} {Deployment} {Applications}.
\newblock page P25.002.
\newblock Conference Name: APS Division of Fluid Dynamics Meeting Abstracts ADS Bibcode: 2021APS..DFDP25002V.

\bibitem[Virtanen et~al., 2020]{virtanen_scipy_2020}
Virtanen, P., Gommers, R., Oliphant, T.~E., Haberland, M., Reddy, T., Cournapeau, D., Burovski, E., Peterson, P., Weckesser, W., Bright, J., van~der Walt, S.~J., Brett, M., Wilson, J., Millman, K.~J., Mayorov, N., Nelson, A. R.~J., Jones, E., Kern, R., Larson, E., Carey, C.~J., Polat, A., Feng, Y., Moore, E.~W., VanderPlas, J., Laxalde, D., Perktold, J., Cimrman, R., Henriksen, I., Quintero, E.~A., Harris, C.~R., Archibald, A.~M., Ribeiro, A.~H., Pedregosa, F., and van Mulbregt, P. (2020).
\newblock {SciPy} 1.0: fundamental algorithms for scientific computing in {Python}.
\newblock {\em Nature Methods}, 17(3):261--272.
\newblock Publisher: Nature Publishing Group.

\bibitem[Walpen et~al., 2023]{walpen_real-scale_2023}
Walpen, A., Catry, G., and Noca, F. (2023).
\newblock Real-{Scale} {Atmospheric} {Wind} and {Turbulence} {Replication} using a {Fan}-{Array} for {Environmental} {Testing} and {UAV}/{AAM} {Validation}.
\newblock In {\em {AIAA} {SCITECH} 2023 {Forum}}. American Institute of Aeronautics and Astronautics.
\newblock \_eprint: https://arc.aiaa.org/doi/pdf/10.2514/6.2023-0812.

\bibitem[Walpen et~al., 2024]{walpen_automated_2024}
Walpen, A., Govoni, T., Stirnemann, J., Ionescu, M., Bosson, N., Catry, G., and Noca, F. (2024).
\newblock Automated control of complex aerodynamic flows generated by {Windshaper} fan arrays.
\newblock In {\em {AIAA} {SCITECH} 2024 {Forum}}, {AIAA} {SciTech} {Forum}. American Institute of Aeronautics and Astronautics.

\bibitem[Wang et~al., 2021]{wang_turbulent_2021}
Wang, N., Kjellberg, C. W.~R., Catry, G., Bosson, N., Sanyal, A., and Glauser, M. (2021).
\newblock Turbulent flows generated by fan array wind tunnel.
\newblock page F20.003.
\newblock Conference Name: APS Division of Fluid Dynamics Meeting Abstracts ADS Bibcode: 2021APS..DFDF20003W.

\bibitem[Wei and Dabiri, 2022]{wei_phase-averaged_2022}
Wei, N.~J. and Dabiri, J.~O. (2022).
\newblock Phase-averaged dynamics of a periodically surging wind turbine.
\newblock {\em Journal of Renewable and Sustainable Energy}, 14(1):013305.

\bibitem[Yos et~al., 2019]{yos_improving_2019}
Yos, D., Gharib, M., and Veismann, M. (2019).
\newblock Improving the {Descent} {Performance} of {Small}-{Scale} {Rotorcraft} through {Added} {Geometries}.
\newblock page NP05.048.
\newblock Conference Name: APS Division of Fluid Dynamics Meeting Abstracts ADS Bibcode: 2019APS..DFDN05048Y.

\end{thebibliography}


\end{document}